# Hardware-In-the-Loop Measurements of the Multi-Carrier Compressed Sensing Multi-User Detection (MCSM) System

Carsten Bockelmann, Fabian Monsees, Matthias Woltering, Armin Dekorsy

Arbeitsbereich Nachrichtentechnik
Universität Bremen
Otto-Hahn-Alle NW1
28359 Bremen
{bockelmann, monsees, woltering, dekorsy}@ant.uni-bremen.de

**Abstract:** MCSM is a recently proposed novel system concept to solve the massive access problem envisioned in future communication systems like 5G and industry 4.0 systems. This work focuses on the practical verification of the theoretical gains that MCSM provides using a Hardware-In-the-Loop (HIL) measurement setup. We present results in two different scenarios: (i) a LoS lab setup and (ii) a non-LoS machine hall. In both scenarios MCSM shows promising performance in terms of the number of supported users and the achieved reliability.

## 1 Introduction

Massive Machine-to-Machine (M2M) communication is expected to be one of the major drivers for new radio access technologies in the development of the 5th generation mobile networks (5G) as well as industrial radio systems for industry 4.0 (I4.0). For cellular networks forecasts indicate that in 2020 a factor 10 to 100 more machine type devices (MTD) than personal mobile phones will be served by a single base station with the number of MTDs connected in the range from 10.000 to 100.000 [NGMN15]. Similarly, the number of devices in industrial contexts will also grow tremendously given the needs of future I4.0 factories. While most current standards like UMTS/LTE or WLAN were conceived for relatively few devices with high data rates, a single MTD often only generates small amounts of user data, shows very diverse channel access or traffic patterns (triggered, periodically, sporadic or random), and for some applications also needs to be low-cost and very energy efficient to operate for long-lifetimes. As a consequence, new radio access technologies are required that are capable of supporting a variety of requirements. On the one hand massive access requires lean signalling and access structures; on the other hand, strict latency or reliability requirements need to be considered. Thus, PHY and MAC technologies with low signalling overhead and adaptable reliability, e.g. in terms of error protection, need to be designed for handling low data rate sporadic machine type communication (MTC) balancing the payload to overhead ratio as well as the reliability.

To this end, we recently proposed the so-called multi-carrier compressed sensing multi-user detection (MCSM) system [MWB15a] which specifically addresses *massive access*, *bandwidth efficiency* and *flexible bandwidth allocation* out of the many challenges in

MTC design. Theoretical work on MCSM has shown promising performance in terms of error rates and the number of connected devices, but practical verification in different scenarios was still missing. The focus of this work is a measurement campaign to show the performance of the proposed MCSM system by a Hardware-In-the-Loop (HIL) setup.

## 2  A multi-carrier system for machine type communication

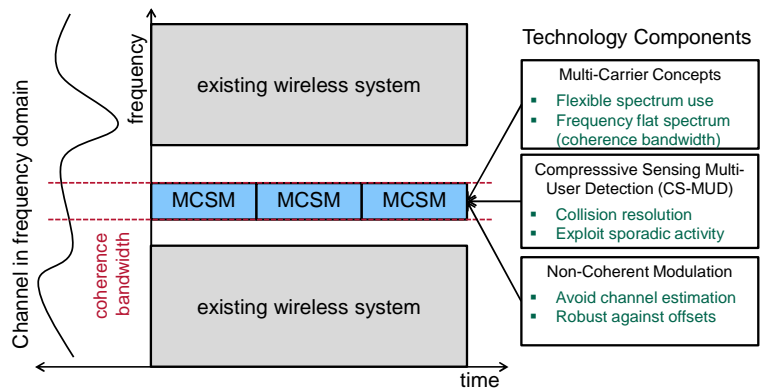

Figure 1 - Multi-Carrier Compressed Sensing Multi-User Detection (MCSM) concept.

Future solutions for either 5G MTC or I4.0 need to fulfil a number of basic requirements that strongly influence the design. First, coexistence has to be guaranteed. In a 5G context coexistence with other services like broadband access is crucial, whereas in I4.0 coexistence with other wireless systems in the same frequency band has to be ensured. Furthermore, we focus on the basic MTC requirements of low signalling overhead, high reliability and flexible resource use with a special attention on large number of served users. The recently proposed MCSM system [MWB15a] tackles these MTC requirements by combining three different technology components as depicted in Figure 1: (i) multi-carrier concepts, (ii) compressive sensing multi-user detection (CS-MUD) and (iii) non-coherent modulation. Each component will be discussed in more detail in the following section.

### 2.1  MCSM Components

*Multi-Carrier-Modulation*

For flexible spectrum allocation multi-carrier systems with carefully designed waveforms have been identified as a potential solution [METI15]. On the one hand, coexistence management is enabled by good spectral containment; on the other hand, spectrally efficient multi-user approaches can be used for low-overhead direct random access communication. Thus, the first technology component of MCSM is a suitable multi-carrier concept to flexibly allocate spectrum for multiple MCSM systems in one frequency band. In this paper we restrict to Orthogonal Frequency Division Multiplexing

(OFDM) as a multi-carrier scheme, but general waveforms providing better spectral containment are equally applicable. Assume that the bandwidth shown in Figure 1 is divided into overall $N$ subcarriers with a subcarrier spacing $\Delta f$. Then a subset of $N_{SC}$ subcarriers of is allocated to one particular MCSM system. MCSM systems are narrowband systems serving up to $K$ nodes per system and several systems can coexist within a certain bandwidth by simply allocating non-overlapping subcarrier blocks. This also enables blanking of sub-bands used by other communication systems. We restrict the following descriptions to a single MCSM system to ease notation. The most important design criterion in choosing $N_{SC}$ determines the bandwidth of the MCSM system: in order to enable non-coherent detection, i.e. the third MCSM component, the bandwidth $N_{SC} \cdot \Delta_f$ has to be smaller than or equal to the coherence bandwidth of the wireless channel $B_c \approx 1/\tau_c$, where $\tau_c$ denotes the delay spread of the channel.

*Compressive Sensing Multiuser Detection (CS-MUD)*

While multi-carrier modulation provides flexible spectrum allocation the physical channel access for each MCSM system has to be designed with very low signalling overhead. To this end, direct random access, i.e. the transmission of payload without a multi-transmission handshake to reserve resources beforehand, has been identified as a crucial component to enable massive access [MWB15a, SBD13a]. The basis for MCSM is CS-MUD, which has been proposed in literature to achieve direct random access with good collision resolution [BSDe13]. CS-MUD is a multi-user detection scheme that exploits certain structures in the multiuser signal. In massive MTC it can be assumed that only a subset of all nodes is active at any time leading to so-called sporadic access of the nodes. This sporadic nature can be exploited at the receiver by modelling inactive nodes as transmitting zeros instead of modulation symbols, yielding an efficient joint detection of activity and user data. The main advantage is a highly reduced signalling overhead compared to known techniques. In this context, sparse multi-user detection facilitating such a joint data detection and signal acquisition is a promising candidate for the application in 5G cellular networks [BSDe13, SBDe13, JSBD14].

*Non-coherent Modulation*

In addition to CS-MUD a third component is required to lower the required signalling overhead even further. Many of today's communication systems employ coherent communication technologies to achieve high data rates at the cost of signalling, i.e. pilots are required for high quality channel estimation. However, differential modulation concepts like Differential M-Phase Shift Keying (D-MPSK) allow for a robust transmission without need for channel estimation as long as the channel can be assumed flat.

## 2.2    MCSM Node Processing and Detection Model

In the following receiver and transmitter side processing are shortly summarized. For a complete theoretical description please refer to [MWB15b].

*Transmitter (Node)*

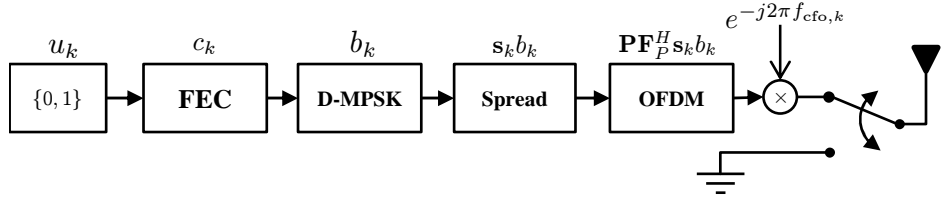

Figure 2 - Transmitter side processing at every MCSM node.

Figure 2 shows the block diagram of a single node out of $K$ nodes summarizing one MCSM system. Each node in a MCSM system can either be active or inactive. In the following, we describe the transmitter processing of an active node.

First, the $N_U$ data bits $\boldsymbol{u}_k \in \{0,1\}^{N_U}$ of a single node $k$ are protected by a forward error correcting code and are modulated according to a D-MPSK modulation. In particular, the encoded information sequence $\boldsymbol{c}_k \in \{0,1\}^{N_C}$ will be modulated to an $M$-Phase Shift Keying (M-PSK) alphabet resulting in symbols $\boldsymbol{a}_k \in \mathbb{C}^L$, where $L$ denotes the frame length. These symbols are differentially modulated to a so called D-MPSK signal through $b_k(i) = a_k(i) b_k(i-1)$, where $i$ denotes the symbol clock and a known starting phase such as $b_k(0) = 1$ is assumed. In the following we restrict ourselves to the case that a MCSM node $k$ spreads each differentially modulated symbol $b_k(i)$ to one OFDM symbol consisting of $N_{SC}$ subcarriers, Figure 1 by using a unique node-specific spreading sequence $\boldsymbol{s}_k \in \mathbb{C}^{N_S}$ that is known at the base station [VER98]. In this case we have $N_S = N_{SC}$ and the symbol clock $i$ remains unchanged for the OFDM symbols. Note that other constellations are possible.

The spread frequency domain symbol is then converted to a time domain symbol via a partial IDFT matrix $\mathbf{F}_P^H \in \mathbb{C}^{N \times N_{SC}}$ such that the node occupies only the $N_{SC}$ out of $N$ subcarriers allocated to that particular MCSM system. Subsequently, a cyclic prefix is added in time domain by multiplication with a matrix $\mathbf{P}$ to counteract inter-symbol interference. Thus, the transmit signal vector $\boldsymbol{z}_k(i)$ of node $k$ in time domain is formally

$$\boldsymbol{z}_k(i) = \boldsymbol{P} \boldsymbol{F}_P^H \boldsymbol{s}_k b_k(i). \qquad (1)$$

*Receiver Processing (Access Point)*

The MCSM detection model employed at the base station contains the symbols of all $K$ nodes in the system that are potentially active and superimposed in time domain. Active as well as inactive users are included in the detection model due to the unknown user activity. Assuming a frequency flat fading channel over the allocated $N_{SC}$ subcarriers (coherence bandwidth), the received signal at the base station for the $i$-th symbol is

$$\mathbf{y}(i) = \sum_{k=1}^{K} \mathbf{F}_P \mathrm{h}_k \mathbf{I} \mathbf{F}_P^H \mathbf{s}_k \mathrm{b}_k(i). \qquad (2)$$

Here, $\mathrm{h}_k \mathbf{I} \in \mathbb{C}^{N_{SC} \times N_{SC}}$ denotes a frequency flat fading channel with a single node specific channel tap. Furthermore, the user symbols $b_k(i)$ can be either D-MPSK or zero which models the inactivity. Clearly, we can reformulate such that $\mathbf{F}_P\mathbf{F}_P^H = \mathbf{I}$, and $\tilde{b}_k(i) = b_k(i)h_k$ is the differentially modulated symbol weighted with the node specific channel coefficient. Hence, we can write the received signal in frequency domain in matrix form as

$$\mathbf{y}(i) = \mathbf{S}\tilde{\boldsymbol{b}}(i), \qquad (3)$$

where $\tilde{\boldsymbol{b}}(i) \in \mathbb{C}^K$ contains the differentially modulated symbols from all $K$ nodes weighted with the individual channel taps. The matrix $\mathbf{S} \in \mathbb{C}^{N_{SC}\times K}$ contains the spreading sequences $\boldsymbol{s}_k$ of all $K$ nodes as column vectors. For a whole transmit frame, of $L$ symbols, we can write

$$\mathbf{Y} = \mathbf{S}\tilde{\boldsymbol{B}}, \qquad (4)$$

where $\mathbf{Y} \in \mathbb{C}^{N_{SC}\times L}$ and $\tilde{\boldsymbol{B}} \in \mathbb{C}^{K\times L}$ summarize the received and transmitted frame, respectively.

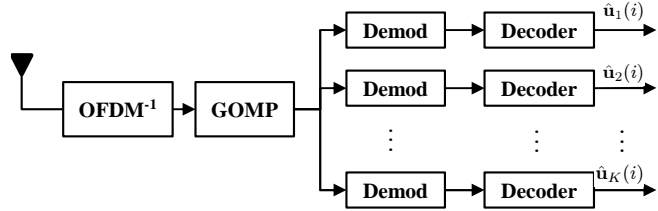

Figure 3 - Base Station processing.

The block diagram of the Base-Station processing is exemplary shown in Figure 3. After converting the time domain symbols to frequency domain leading to (4), a Group Orthogonal Matching Pursuit (GOMP) is employed as Compressed Sensing algorithm to detect the activity of the nodes. The GOMP performs two steps consecutively: first it estimates the activity of one node and, subsequently, estimates the data of that node. This procedure is repeated until a certain stopping criterion is met. For a more detailed description of the GOMP the reader is referred to [SBD13b]. At the output of the GOMP, we have estimates $\hat{b}_k(i)$ for the modulated symbols of the active nodes. The receiver recovers the modulation symbols by performing differential demodulation

$$\hat{a}_k(i) = \frac{\hat{b}_k(i)}{\hat{b}_k(i-1)} = \hat{b}_k(i)\hat{b}_k(i-1)^*, \qquad (5)$$

which are subsequently demapped to the code word estimate $\hat{\boldsymbol{c}}_k$. Finally, the estimated information bit sequence $\hat{\boldsymbol{u}}_k$ is obtained by decoding the code word. Please note that this is the simplest form of a D-MPSK demodulator and more advanced techniques exist but are out of the scope of this paper.

## 3 Measurement Setup

A practical measurement with a simplified transmission scenario is implemented to evaluate the proposed MCSM system. Two development hardware platforms serve as measurement environment. First, the platforms, the frame structure and the modelling of the massive access will be introduced. Then, we shortly describe the two different measurement scenarios: 1) A Line of Sight (LoS) setup within our laboratory and 2) a non-Line of Sight (non-LoS) setup in a machine hall, where different interference factors like metal switch cabinets, machines and material affect the transmission.

### 3.1 Hardware Platform and Setup

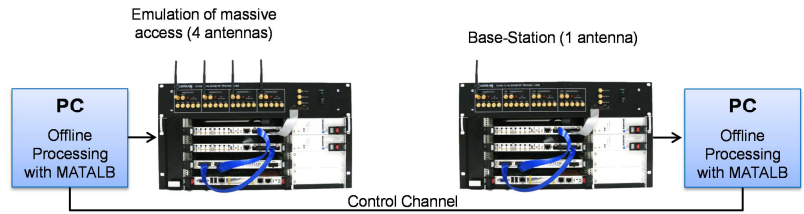

Figure 4 - Hardware-in-the-Loop (HIL) setup.

Both Nutaq hardware development platforms used in our measurement setup are equipped with an 8 channel 14 Bit AD converter with up to 125 MSPS as well as an 8 channel 14 Bit DA converter with up to 500 MSPS. The maximum sampling rate is 104 MHz, which is decreased to 26 MHz for this measurement due to memory bandwidth limitations. Hence, a sample timing of $T_s = 1/26\ MHz = 38.462\ ns$ results.

Furthermore, both provide transceivers with four RF Frontends with 26 dBm maximum transmission power for MIMO transmission within the 2.4 GHz and 5 GHz ISM Bands. The RF frontend is implemented as a non-DC-coupled highpass, where all frequencies larger 100 Hz up to a maximum bandwidth of 20 MHz pass the frontend. The typical CFO is specified as 100ppm of the actual carrier frequency. Assuming an ISM Band with a carrier frequency $f_{lo} = 2.4\ GHz$ a CFO of approx. $f_{CFO,i} \approx \pm 2.4\ kHz$ per transceiver is expected. During operation CFOs of up to $f_{CFO} = (f_{lo,TX} + f_{CFO,TX}) - (f_{lo,RX} + f_{CFO,RX}) \approx 5\ kHz$ can be observed, which is well within specifications. Furthermore, both platforms contain a Windows PC that controls the hardware environment and facilitates the offline processing of the HIL setup using MATLAB. A direct Ethernet connection between both platforms simulates an ideal feedback control channel to enable automated measurements and error rate calculations.

### 3.2 Frame Structure and Emulation Massive Machine Access

The practical evaluation of massive access faces the challenge to realize a massive number of nodes that access a common base station. As illustrated in Figure 4 the two available transceivers are setup as HIL-devices: one serve as base station (RX), the other servers as a "user emulator" (TX). The base station processing is relatively straightforward along the lines described in section 2.2 with additional steps to ensure synchronization, CFO estimation and so on. The "user emulation", however, requires the generation of many virtual users and their mapping to the available hardware, i.e. primarily the four available antennas to emulate channel variations.

*Frame Structure and Node Parameters*

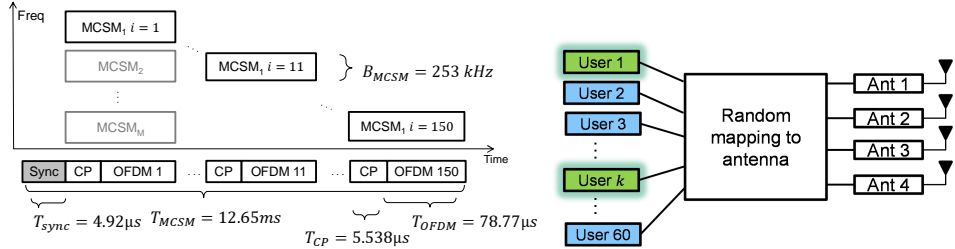

Figure 5 - Frame design and symbol timing (left) and virtual massive access (right)

The overall frame structure of our MCSM system is depicted in Figure 5. Multiple MCSM systems occupy the available 20 MHz bandwidth at a carrier frequency of 2.484 GHz. As a starting point to show the massive access capabilities of MCSM we target an LTE like implementation. The left part of Table 1 summarizes the resulting system parameters in comparison with LTE parameters. Additionally, Figure 5 (left) shows the principle timing of one OFDM symbol: the bandwidth is subdivided by an $N = 2048$ point-IFFT leading to a core symbol timing is $T_{OFDM} = 2048/26 MHz = 78.77\ \mu s$ with a sub-carrier spacing $\Delta f = 1/T_{OFDM} = 12.695\ kHz$. Additionally, the cyclic prefix (CP) of length 144 adds $T_{CP} = 5.538\ \mu s$ to the overall OFDM symbol.

| Parameter | Var | MCSM | LTE | Parameter | Var | MCSM |
|---|---|---|---|---|---|---|
| **Carrier Frequency** | $f_{lo}$ | $2.484\ GHz$ | See 3GPP LTE | **# of users in on MCSM block** | $K$ | 60 |
| **Sampling time** | $T_s$ | $38.462\ ns$ | $32.552\ ns$ | **# info bits per user** | $N_u$ | 150 |
| **IFFT Length** | $N$ | 2048 | 2048 | **Channel Code** | | [5;7] Conv. Code |
| **OFDM Symbol time** | $T_{OFDM}$ | 78.77 µs | 66.6666 µs | **Modulation** | | 4-DPSK |
| **CP Length** | $N_{CP}$ | 144 | 144/160 or 512 | **# spreading sequence length** | $N_s$ | 20 |
| **CP Symbol time** | $T_{CP}$ | 5.538 µs | 4.6875 µs | **Occupied bandwidth** | $B_{MCSM}$ | $253\ kHz$ |
| **Subcarrier spacing** | $\Delta f$ | $12.69\ kHz$ | $15\ kHz$ | **# OFDM symbols** | $N_{OFDM}$ | 150 |

Table 1 – System parameters: OFDM compared to LTE (left) and node parameters (right).

Each MCSM system and the emulated nodes are parametrized according to the right part of Table 1. Each node transmits $N_U$ information bits encoded by a half rate convolutional encoder with generators [5;7]. Due to the spreading a single symbol is transmitted per OFDM symbol such that one MCSM block occupies an overall bandwidth of $B_{MCSM} =$

$N_{SC}\Delta f = 253\ kHz$ and consists of 150 OFDM symbols. This gives a total frame length of $N_{MCSM} = N_B * (N + N_{CP}) = 328800$ sample points and a total frame duration of $T_{Frame,MCSM} = N_{MCSM} * T_s = 12.65\ ms$, which is slightly longer than the LTE frame of $T_{Frame,LTE} = 10\ ms$. Additionally, a regular frequency hopping of the MCSM systems is applied every 10 OFDM symbol to exploit frequency diversity if the channel is not frequency flat or changing over time. Note that more general frequency hopping strategies can also be applied. To evaluate the start of one frame an LTE related synchronization sequence of 128 samples or $T_{sync} = 4.92\ \mu s$ is added to the frame. The virtual massive access of all nodes is realized at the transmitter (TX) seen in Figure 5 (right). Here, all signals of every active node (highlighted) are generated like described in (2). Each active node is randomly mapped to one antenna to emulate different user channels. The inactive nodes are modelled to be zero and therefore are not mapped to an antenna.

### 3.3  Measurement Environments

*Setup 1: Line of Sight (LoS)*

Figure 6 shows the setup of the LoS scenario. The transmitter simulating the massive access (TX) is in the upper left, the receiver (RX) is located roughly 5 m away close to the door. The Line of Sight is unobstructed and mostly wooden furniture (some tables, cabinets, etc.) may act as scatters. Hence, the scenario is expected to show a mostly frequency flat behaviour which is reflected by the channel measurements depicted on the right.

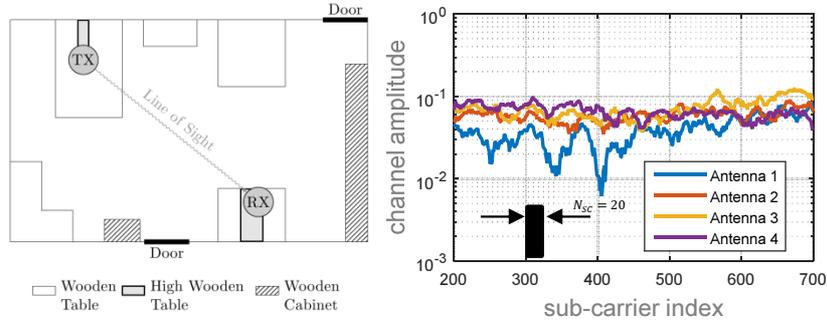

Figure 6 - LoS scenario: (left) Schematic of the laboratory setup; (right) exemplary channel measurement with subcarrier 200 up to 700 and the bandwidth of the MCSM system considered.

*Setup 2: Non-Line of Sight (non-LoS)*

As a second setup we choose a more realistic industrial environment, i.e. a machine hall used for automation laboratories and research. The left part of Figure 7 shows a picture of the scenario. The TX and the RX are located as far away as possible, which leads to a larger distance of roughly 14.5 m. Furthermore, the line of sight is now obstructed by metal switch cabinets, metal machines and other electrical components that are located within this hall.

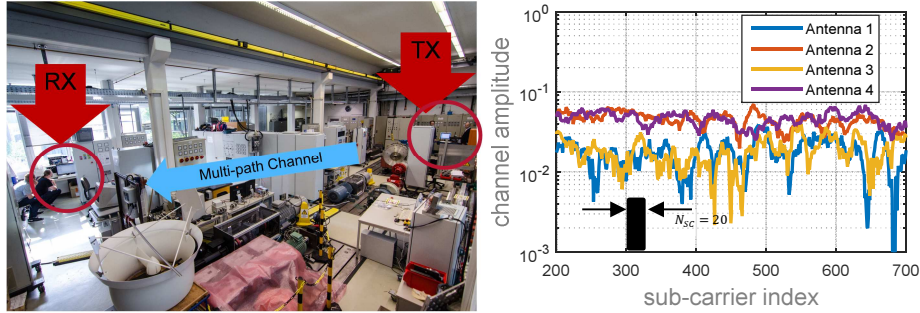

Figure 7 – Non-LoS scenario – (left) Schematic of the machine hall setup; (right) exemplary channel measurement.

In addition to the setup 1 with only wooden cabinets and tables a high degree of reflections and stronger dampening of the signal is to be expected. Hence, the measured channel seen on the right of Figure 7 is much more frequency selective.

## 4 Measurement Results

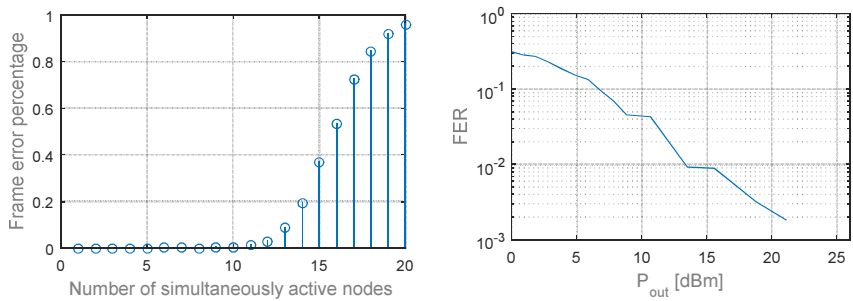

Figure 8 - Measured frame error percentage vs. the number of active users in LoS setup (left). Frame Error Rate vs. transmit power in non-LoS setup with random node activity (right).

Figure 8 shows the results of our measurements. The left plot depicts the frame error percentage for different node activities at a fixed transmit power in the LoS case (setup 1). We see that MCSM performs highly reliable if up to 10 nodes are simultaneously active. Increasing the number of active nodes gracefully increases the likelihood for frame errors due to the higher system load. In setup 1 MCSM is mainly limited by the node activity due to the frequency flat channel and sufficiently high received power. To support higher number of active nodes, the MCSM parametrization can be adapted towards longer spreading codes and higher bandwidth allocation

In contrast, the right plot of Figure 8 shows the average frame error rate (FER) performance of MCSM in a non-LoS scenario (setup 2) with random node activity. The nodes are active with probability $p_a = 0.1$, which means we have $0.1 * 60 = 6$ simultaneously active nodes on average using 273 kHz in a single MCSM system. In setup 2 the influence of different transmit powers can be observed due to the multipath channel and the additional path loss compared to the LoS setup. For low transmit powers the FER is severely degraded but for high transmit powers a comparable performance to the LoS setup is achievable. For maximum transmit power of 26 dBm no errors could be observed during our measurement. Naturally, the coverage of any MCSM system given a minimum required reliability is strongly coupled with the propagation environment and allowed transmit powers.

**Conclusion**

In this work we have presented a practical evaluation of the MCSM system concept in a lab environment as well as a more realistic machine hall setup. Our measurements validate the theoretical findings of previous works with respect to the robustness and flexibility of the scheme. The presented results are only one exemplary MCSM parametrization which is specifically tuned to support a high number of potential users in massive M2M situations. Furthermore, other parametrizations can be tuned towards latency reduction (frame length) and higher reliability, e.g. by longer spreading sequences or by using stronger codes. In conclusion we have shown that the recently introduced MCSM system can be considered as a candidate technology for novel MTC applications in 5G and I4.0 applications.

## 5   References

[BSD13]     Bockelmann, C.; Schepker, H.; Dekorsy, A.: Compressive Sensing based Multi-User Detection for Machine-to-Machine. In Transactions on Emerging Telecommunications Technologies: Special Issue on Machine-to-Machine: An emerging communication paradigm, Vol. 24, No. 4, pp. 389-400, June 2013.

[SBD13a]    Schepker, H.; Bockelmann, C.; Dekorsy, A.: Exploiting Sparsity in Channel and Data Estimation for Sporadic Multi-User Communication. In: Proc. 10th International Symposium on Wireless Communication Systems (ISWCS 13), Ilmenau, Germany, 27. - 30. August 2013

[SBD13b]    Schepker, H.; Bockelmann, C.; Dekorsy, A.: Improving Group Orthogonal Matching Pursuit with Iterative Feedback. In: Proc. IEEE 78th Vehicular Technology Conference (VTC 2013-Fall), Las-Vegas, USA, 2. - 5. September 2013

[JSB14]     Ji, Y.; Stefanović, Č.; Bockelmann, C.; Dekorsy, A.; Popovski, P.: Characterization of Coded Random Access with Compressive Sensing based Multi-User Detection. In: Proc. IEEE Globecom 2014, Austin, TX, USA, 8. - 11. December 2014

[NGMN15]    NGMN: 5G White Paper, 2015

[METI15]    METIS D6.6: Final Report on the METIS System Concept and Technology Roadmap, 2015.

[MWB15a]    Monsees, F.; Woltering, M.; Bockelmann, C.; Dekorsy, A.: Compressive Sensing Multi-User Detection for Multi-Carrier Systems in Sporadic Machine Type Communication. In: Proc. IEEE 81th Vehicular Technology Conference (VTC2015-Spring), Glasgow, Great Britain, May 2015.

[VER98]     S. Verdu, Multiuser Detection, Cambridge University Press, 1998

[MWB15b]    Monsees, F.; Woltering, M.; Bockelmann, C.; Dekorsy, A.: A Potential Solution for MTC: Multi-Carrier Compressive Sensing Multi-User Detection. In: Proc. IEEE The Asilomar Conference on Signals, Systems, and Computers, Asilomar Hotel and Conference Grounds, USA, November 2015.